\documentclass[11pt,twoside]{article}
\usepackage{./asp2010}

\usepackage{graphicx}

\resetcounters

\begin{document}

\title{Magnetospheres of massive stars across the EM spectrum}
\author{V. Petit$^1$, S.~P. Owocki$^2$, M.~E. Oksala$^2$ and the MiMeS Collaboration
\affil{$^1$Dept. of Geology \& Astronomy, West Chester Univ., West Chester PA, USA}
\affil{$^2$Dept. of Physics, University of Delaware, Newark DE 19716, USA}}

\begin{abstract}
Magnetic massive stars -- which are being discovered with increasing frequency -- represent a new category of wind-shaping mechanism for O and B stars. Magnetic channeling of these stars' radiation-driven winds, the Magnetically Confined Wind Shock paradigm, leads to the formation of a shock-heated magnetosphere, which can radiate X-rays, modify UV resonance lines, and create disks of H$\alpha$ emitting material. The dynamical properties of these magnetospheres are well understood from a theoretical point of view as an interplay between the magnetic wind confinement and rotation. However, the manifestations of magnetospheres across the spectrum may be more complex and various than first anticipated. On the other hand, recent advances in  modeling these magnetospheres provide a key to better understand massive star winds in general. We will summarize the coordinated observational, theoretical, and modeling efforts from the Magnetism in Massive Star Project, addressing key outstanding questions regarding magnetosphere manifestations across the spectral domain.
\end{abstract}

\section{Magnetosphere 101}

Stellar magnetospheres form through the channelling and confinement of an outflowing stellar wind by the star's magnetic field, like the large-scale, often nearly dipolar fields found in massive stars. This idea of a magnetosphere has been exploited to explain particular properties of some massive stars, for example the photometric light curve and H$\alpha$ variations of the He-Strong star $\sigma$\,Ori\,E \citep{VP_1978ApJ...224L...5L}, the UV variations of magnetic Bp stars \citep{VP_1990ApJ...365..665S} and the X-rays properties of the O-type star $\theta^1$\,Ori\,C \citep{VP_2005ApJ...628..986G}.  

The global competition between the magnetic field and stellar wind can be characterized by the so-called wind magnetic confinement parameter $\eta_\star \equiv B^2_\mathrm{eq}R^2_\star / \dot{M}\varv_\infty$, which depends on the star's equatorial field strength ($B_\mathrm{eq}$), stellar radius ($R_\star$), and wind momentum ($\dot{M}\varv_\infty$). For a dipolar field, one can identify an Alfv\'en radius $R_\mathrm{A}\equiv\eta_\star^{1/4}R_\star$, representing the extent of strong magnetic confinement. 
Above $R_\mathrm{A}$, the wind dominates and stretches open all field lines. But below $R_\mathrm{A}$, the wind material is trapped by closed field line loops, and in the absence of significant stellar rotation is pulled by gravity back onto the star within a dynamical (free-fall) time-scale \citep{VP_2002ApJ...576..413U}.

In the presence of significant stellar rotation, centrifugal forces can support any trapped material above a Kepler co-rotation radius, $R_\mathrm{K}\equiv (\mathrm{v}_\mathrm{rot}/\mathrm{v}_\mathrm{crit})^{-2/3}$, where $\mathrm{v}_\mathrm{rot}$ is the surface rotation speed, and $\mathrm{v}_\mathrm{crit}$ is the critical speed at which surface material would be in Keplerian orbit \citep{VP_2008MNRAS.385...97U}. This requires that the magnetic confinement extend beyond this Kepler radius, i.e. $R_\mathrm{A}>R_\mathrm{K} $, in which case the rotationally supported, magnetically confined material accumulates to form a ``\textit{centrifugal magnetosphere}'' (CM). In the opposite limit of $R_\mathrm{K}>R_\mathrm{A}>R_\star$, representing strong magnetic confinement but slow rotation, the transient suspension of trapped material in closed loops establishes a ``\textit{dynamical magnetosphere}'' (DM).   

\begin{figure}[t]
\begin{center}
\includegraphics[width=5.00in]{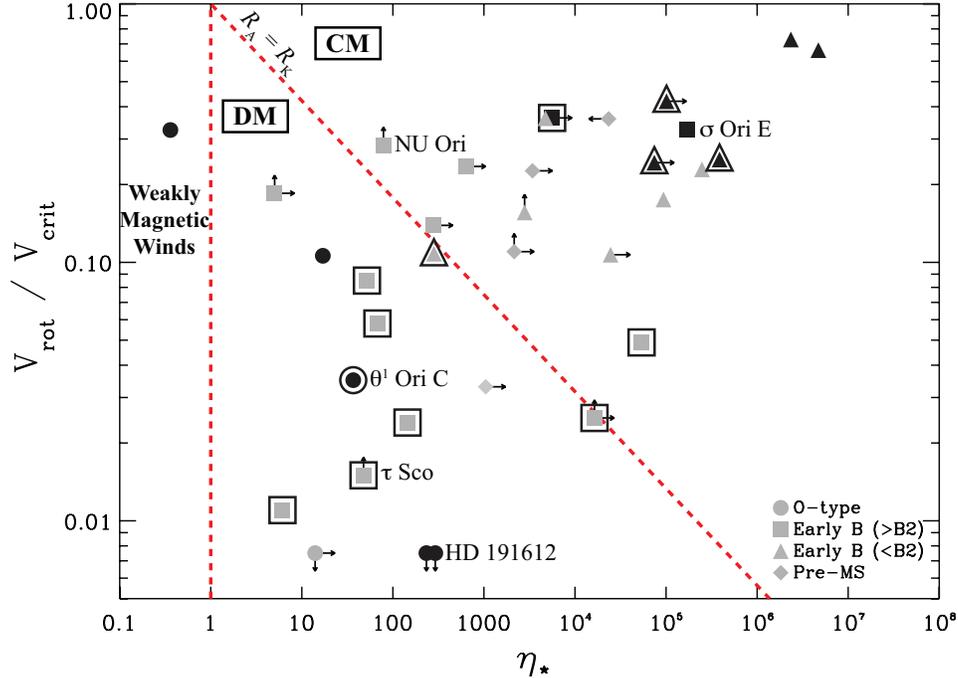}
\caption{\label{ref_VPetit_fig1}
Parameter plane representing rotation ratio versus magnetic confinement. The boundary defined by $R_\mathrm{A}=R_\mathrm{K}$ and the separation between strong confinement and weakly magnetized winds ($\eta_\star=1$) are indicated by dashed lines. 
The arrows indicate limits when only $v\sin i$ measurements and/or longitudinal field measurements are available
 instead of periods and dipolar field strengths. The $\eta_\star$ were calculated using the CAK theory wind momentum. The darker symbols indicate stars for which rotationally modulated H$\alpha$ emission is present. The encircled symbols represent stars displaying rotational modulation of their UV resonance lines. }
\end{center}
\end{figure}

As illustrated in Fig. \ref{ref_VPetit_fig1}, in a parameter plane of rotation ratio ($\mathrm{v}_\mathrm{rot}/\mathrm{v}_\mathrm{crit}$) versus magnetic confinement ($\eta_\star$), the boundary defined by $R_\mathrm{A}=R_\mathrm{K}$ separates the domain between DM (lower left) and CM (upper right). The vertical line at $\eta_\star=1$ separates the domain for strong confinement and magnetosphere formation (right) and weakly magnetized winds (left), for which all magnetic field lines become opened by the outflowing wind. The symbols represent types of stars, as described in Sec. 2.

The first attempts to model magnetized winds and resulting magnetospheres of massive stars used direct 2D MHD simulations applied to the case of slow rotation or field aligned rotation, with confinement parameters restricted to $\eta_\star<10^3$ \citep{VP_2002ApJ...576..413U,VP_2003PhDT.........1U,VP_2008MNRAS.385...97U}.  But some magnetic massive stars, particularly the strongly magnetic Bp stars with weak winds (e.g. $\sigma$\,Ori\,E), are inferred to have $\eta_\star>10^5$. For a typical rotation of half critical, the magnetic field is strong enough to justify a \textit{Rigidly Rotating Magnetosphere} \citep[RRM,][]{VP_2005MNRAS.357..251T} approach to describe the centrifugally supported circumstellar material. A subsequent \textit{Rigid Field Hydro-Dynamics} \citep[RFHD,][]{VP_2007MNRAS.382..139T} model follows the dynamical channeling of the wind material into the magnetosphere. Both RRM and RFHD methods have the advantage that they can be generalized to the fully 3D cases like a tilted dipole, or even arbitrary field topology. Full 3D MHD is a challenging task, but preliminary results look promising (see contribution from ud-Doula et al., these proceedings).

\section{Magnetosphere diagnostics}

As an effort to explore the properties of massive star magnetospheres, the Magnetism in Massive Stars (MiMeS) Collaboration \citep[][see also Grunhut et al. these proceedings]{VP_2010arXiv1009.3563W} has been collecting spectropolarimetric observations of massive stars in order to first expand the population of known magnetic massive stars and secondly to provide modern determination of their magnetic field characteristics. 
We have used the MiMeS extensive database of magnetic stars to populate the rotation ratio versus confinement diagram with a sample of magnetic massive OB stars (data points in Fig. \ref{ref_VPetit_fig1}). The circles represent O-type stars, while the diamonds show pre-main-sequence Herbig stars. We have separated the early B-type star sample in two, with a cutoff at $T=22\,000$\,K which roughly represents spectral type B2. 
In this section, we provide a guided-tour of the principal magnetospheric diagnostics seen in massive stars.  The specific stars mentioned throughout this discussion are labeled in Fig. \ref{ref_VPetit_fig1}.

 \subsection{Optical and Ultraviolet}

The He-strong B-type star $\sigma$\,Ori\,E is well known for its doubled-peaked H$\alpha$ emission, which is modulated over its short 1.2\,d rotational period \citep{VP_1974ApJ...191L..95W}. The strong dipolar magnetic field ($\sim10$\,kG) and the rapid rotation place $\sigma$\,Ori\,E in the upper right corner of the diagram, well into the CM domain.  
As the magnetic field is tilted with respect to the rotation axis, the RRM model predicts that the material accumulates primarily at the intersection of the magnetic and rotational equators, forming two azimuthal ``clouds''. The model reproduces well the observed H$\alpha$ rotational modulation of $\sigma$\,Ori\,E \citep{VP_2005ApJ...630L..81T}.
Furthermore, the transit of these clouds in front of the star provides an interpretation to the eclipses seen in the optical light curve of $\sigma$\,Ori\,E, as well as other Bp stars \citep{VP_2008MNRAS.389..559T}. 

The darker symbols in Fig. \ref{ref_VPetit_fig1} represent stars for which such modulated H$\alpha$ emission is observed. There is a cluster of B-type stars, located in the CM domain, which display magnetospheric characteristics similar to $\sigma$\,Ori\,E, although the exact emission pattern varies with the different field geometries \citep[e.g.][]{VP_2010MNRAS.405L..51O,VP_2011AJ....141..169B,VP_2011arXiv1109.3157G}. 

Note also that all the O-type stars in the DM region display modulated H$\alpha$ emission, while none of the B-type stars in this region do. 
Even if the residence time of individual parcels of material in the closed loops is short, a global overdensity of plasma streams develops near the magnetic equator. A large source of mass, such as the one associated with the strong wind of these O-type stars, would therefore be needed to produce a detectable H$\alpha$ emission. For example, MHD simulations can reproduce the observed H$\alpha$ rotational modulation of the magnetic O-type star HD\,191612  (Sundqvist et al. in prep).


Given their low mass-loss rate, the magnetosphere of B-type stars located in the DM domain are not expected to show significant H$\alpha$ emission, as they cannot accumulate wind material over time like the B-type stars in the CM domain. However, another way to probe these magnetospheres is through their UV properties. 
A main example is the early B-type star $\tau$\,Sco, located in the DM domain, for which the C\textsc{iv}\,$\lambda\lambda$1548,~50 lines vary with the 40\,d rotational period \citep{VP_2006MNRAS.370..629D}, and display superionization of the N\textsc{v} doublet \citep[see][]{VP_2011MNRAS.412L..45P,VP_2011MNRAS.416.1456O}. 
A number of new magnetic stars have been discovered based on such UV proxies \citep[e.g.][]{VP_2003A&A...411..565N,VP_2011MNRAS.412L..45P,VP_2011IAUS..272..192H}.

The encircled symbols in Fig. \ref{ref_VPetit_fig1} represent stars for which rotational modulation of UV resonance lines is observed. 
Clearly, the presence of the magnetic field exerts an influence on the ionization balance of the stellar winds in both DM and CM domains and should provide a robust diagnostic for magnetism. Although some phenomenological understanding is available \citep[e.g.][]{VP_1990ApJ...365..665S}, quantitative models are still being developed \citep{VP_2007ASPC..361..488S}.

\subsection{X-rays}

Massive stars are generally X-ray bright due to the intrinsic instability of the line-driving mechanism of the radiative stellar winds \citep{VP_1997A&A...322..878F,VP_2001ApJ...559.1108O}. The magnetically channeled wind shocks (MCWS scenario) associated with magnetic massive stars should also lead to copious X-ray emission, by the radiative cooling of the shock heated plasma \citep{VP_1997ApJ...485L..29B}. 
For example, the X-rays from the O-type star $\theta^1$\,Ori\,C are more luminous and harder than in typical O-stars, and modulated by the rotation period. \citet{VP_2005ApJ...628..986G} used 2D MHD simulations, including an explicit energy equation, to track the shock heated material and its radiative cooling, and were able to reproduce the X-ray properties of $\theta^1$\,Ori\,C. Therefore, it seems at first glance that luminous, hard and variable X-ray emission could be a proxy for magnetism in massive stars. For example, the B-type star $\tau$\,Sco also displays a hard X-ray spectrum \citep{VP_1994ApJ...421..705C}. 
However, as counter example, the O-type star HD\,191612 is quite luminous, but has a rather soft spectrum \citep{VP_2010A&A...520A..59N}. The same is true for the B-type star NU\,Ori, which is not only soft, but does not show any significant variation over the duration of a $\sim10$\,d \textit{Chandra} observation \citep{VP_2005ApJS..160..557S}. 

While no clear empirical picture has emerged yet, X-rays provide a different perspective on the important question of massive star wind mass-loss rate. 
For example, RFHD simulations show that the overall X-ray flux of stars in the CM region is quite sensitive to the mass loss rate. The distribution in temperature of the differential emission measure (DEM) is governed by both the pre-shock and post-shock characteristics of the magnetosphere. Both of these are affected by the wind properties, with the post-shock cooling length being longer for lower-density wind flows. The RFHD model predicts a DEM distribution that can, for example, be confronted to that derived from \textit{Chandra} ACIS observations of magnetic massive stars \citep{VP_2011IAUS..272..194H}.

\section{Conclusion}
In conclusion, magnetic confinement and rotation are two main ingredients that dictates the structure of a magnetosphere for a given magnetic field topology. From these ingredients, we can identify two main magnetosphere regimes (DM, CM).

We understand quite well H$\alpha$ emission from both DM and CM, and the codes are now ripe for more quantitative modeling and predictions.

UV is a good proxy for magnetism, even for stars in DM with too low mass-rate to display H$\alpha$ magnetospheric emission, but detailed models are still being developed.

X-rays arising from MCWS provide a promising avenue for probing the physics of magnetospheres not accessible through other diagnostics, but the correspondence to specific stars is not fully understood.

\acknowledgements
We would like to thank D.~A. Bohlender, D.~H. Cohen, M. Gagn\'e, H.~F. Henrichs, Th. Rivinius, J.~O. Sundqvist, R.~H.~D. Townsend, A. ud-Doula and G.~A. Wade.

\bibliography{VPetit}

\end{document}